\documentclass[%
reprint,
superscriptaddress,
amsmath,
amssymb,
aps,
pra,
floatfix,
]{revtex4-2}
\usepackage{graphicx}
\usepackage{dcolumn}
\usepackage{bm}
\usepackage[utf8]{inputenc}
\usepackage[T1]{fontenc}
\usepackage{mathptmx}
\usepackage{textcomp}
\usepackage{siunitx}
\usepackage{bigints}

\usepackage{hyperref}
\usepackage[switch]{lineno}
\usepackage[table,xcdraw,dvipsnames]{xcolor}
\usepackage{booktabs}
\usepackage{float}
\usepackage{color}
\usepackage{ulem}
\usepackage{multirow}

\begin{document}
\title[Single device offset-free magnetic field sensing principle with tunable sensitivity and linear range based on spin-orbit-torques]{Single device offset-free magnetic field sensing principle with tunable sensitivity and linear range based on spin-orbit-torques}

\author{Sabri~Koraltan}
\email{sabri.koraltan@univie.ac.at}
\affiliation{Physics of Functional Materials, Faculty of Physics, University of Vienna, Vienna, Austria}%
\affiliation{Vienna Doctoral School in Physics, University of Vienna, Vienna, Austria}%
\author{Christin~Schmitt}
\affiliation{Institute of Physics, Johannes Gutenberg University Mainz, 55099 Mainz, Germany}%

\author{Florian~Bruckner}
\affiliation{Physics of Functional Materials, Faculty of Physics, University of Vienna, Vienna, Austria}%
\author{Claas~Abert}%
\affiliation{Physics of Functional Materials, Faculty of Physics, University of Vienna, Vienna, Austria}%
\affiliation{Research Platform MMM Mathematics-Magnetism-Materials, University of Vienna, Vienna, Austria.}%
\author{Klemens~Prügl}%
\affiliation{Infineon Technologies AG, Regensburg, Germany}
\author{Michael~Kirsch}%
\affiliation{Infineon Technologies AG, Regensburg, Germany}
\author{Rahul~Gupta}
\affiliation{Institute of Physics, Johannes Gutenberg University Mainz, 55099 Mainz, Germany}%
\author{Sebastian~Zeilinger}
\affiliation{Physics of Functional Materials, Faculty of Physics, University of Vienna, Vienna, Austria}%
\affiliation{Infineon Technologies AG, Villach, Austria}
\author{Joshua~M.~Salazar-Mej\'ia}
\affiliation{Physics of Functional Materials, Faculty of Physics, University of Vienna, Vienna, Austria}%
\affiliation{Vienna Doctoral School in Physics, University of Vienna, Vienna, Austria}%
\author{Milan~Agrawal}
\affiliation{Infineon Technologies AG, Munich, Germany}
\author{Johannes~G\"{u}ttinger}
\affiliation{Infineon Technologies AG, Villach, Austria}
\author{Armin~Satz}
\affiliation{Infineon Technologies AG, Villach, Austria}
\author{Gerhard~Jakob}
\affiliation{Institute of Physics, Johannes Gutenberg University Mainz, 55099 Mainz, Germany}%
\author{Mathias~Kl\"{a}ui}
\affiliation{Institute of Physics, Johannes Gutenberg University Mainz, 55099 Mainz, Germany}%

\author{Dieter~Suess}%
\affiliation{Physics of Functional Materials, Faculty of Physics, University of Vienna, Vienna, Austria}%
\affiliation{Research Platform MMM Mathematics-Magnetism-Materials, University of Vienna, Vienna, Austria.}%

\date{\today}

\begin{abstract}
We propose a novel device concept using spin-orbit-torques to realize a magnetic field sensor, where we eliminate the sensor offset using a differential measurement concept. We derive a simple analytical formulation for the sensor signal and demonstrate its validity with numerical investigations using macrospin simulations. The sensitivity and the measurable linear sensing range in the proposed concept can be tuned by either varying the effective magnetic anisotropy or by varying the magnitude of the injected currents. We show that undesired perturbation fields normal to the sensitive direction preserve the zero-offset property and only slightly modulate the sensitivity of the proposed sensor. Higher-harmonics voltage analysis on a Hall cross experimentally confirms the linearity and tunability via current strength. Additionally, the sensor exhibits a non-vanishing offset in the experiment which we attribute to the anomalous Nernst effect.
\end{abstract}
\maketitle
\section{Introduction}

Spintronics is is a very active field of research since the discovery of the giant magnetoresistive (GMR) effect\cite{baibich_giant_1988,binasch_enhanced_1989}.
With the ability to control the magnetization of magnetic materials using spin-polarized currents\cite{zhang_mechanisms_2002, slonczewski_current-driven_1996,manchon_current-induced_2019}, new possibilities have emerged for spintronic devices\cite{bader_spintronics_2010, bhatti_spintronics_2017, shao_roadmapSOT_2021, chumak_roadmap_2022}.
Spin-orbit-torque (SOT) induced magnetization switching processes have proven to be ultra-fast and energy efficient\cite{miron_perpendicular_2011,garello_ultrafast_2014,cubukcu_spin-orbit_2014,zhang_spin-orbit_2015, avci_current-induced_2017}. Additionally, spin currents have been used to propagate chiral objects such as domain walls\cite{miron_fast_2011}, or skyrmions\cite{jiang_skyrmions_2017,everschor_perspective_2018,kern_deterministic_2022,velez_current-driven_2022}.
A newly developing topic in this area is the usage of SOT for the magnetic field sensing\cite{braganca_nanoscale_2010, shao_roadmapSOT_2021, xu_ultrathin_2018, li_spinorbit_2021, suess_device_2021}.

Accurate sensors for magnetic field measurements\cite{heremans_solid_1993} are very important for various applications like biosensors and bio-chips for medical usage\cite{graham_magnetoresistive-based_2004}, read heads for magnetic recording devices\cite{mcfadyen_state---art_2006}, or for position measurements with the vortex sensors used in the automotive industry \cite{suess_topologically_2018}.
Conventional magnetoresistive field sensors rely on the response of a rather soft magnetic layer to the external field. An additional hard magnetic layer that is insensitive to changes of the external field acts as a reference system introducing a field-dependent magnetoresistivity. The linear regime for these systems can be tuned to various field ranges by adjusting the material properties of the soft magnetic layer.

In order to obtain more accurate sensing signals, the conventional magnetoresistive sensors are operated in Wheatstone bridge configurations, leading to further problems such as fabrication tolerances of different elements, thermal drift within and in between different elements.
All magnetoresistive measurements exhibit a positive onset due to Ohm's law, which makes it difficult to get accurate measurement results for small external fields, known as the sensing offset. While the reduction of this offset leads to an improvement in the sensing process, it can enhance the resolution of the device to determine vanishing or very small fields.
Reducing the offset present at vanishing fields is a key challenge in magnetic-field measurement applications. Several designs exist where the offset can be reduced, or nearly eliminated. These solutions are either based on spinning current techniques for Hall sensors, or still use Wheatstone bridges, where four distinct magnetic stacks are required to eliminate the offset \cite{xu_ultrathin_2018}. Recently, a 3D magnetic field sensor was reported using SOTs \cite{li_spinorbit_2021}, where the change in domain wall positions leads to a sensor signal. However, in both cases the linear range is extremely small.  

In this work, we propose a sensor concept for magnetic-field measurements that exploits the symmetric nature of SOT in order to eliminate the sensing offset. Using micromagnetic simulations we demonstrate that a differential measurement method using SOT leads to the elimination of the offset error.
We reveal how increasing the amplitude of the injected current can be used to tune the sensitivity of the sensor, whereas higher perpendicular magnetic anisotropies allow for enhancing the linear range. The robustness of the measurement scheme is tested by investigating the role of perturbation fields normal to the sensitive direction on the offset, and sensitivity.
The proposed concept is experimentally tested on a Hall-cross stack, where a Ta layer is used as the heavy-metal layer and spin-polarizer, and a thin film of CoFeB is deposited on top as the magnetic layer. Second harmonics measurements are performed to obtain the linear sensor signal via the anomalous Hall effect.

\section{\label{sec:level2}CONCEPT}
\begin{figure*}
\centering
\includegraphics[width=\textwidth]{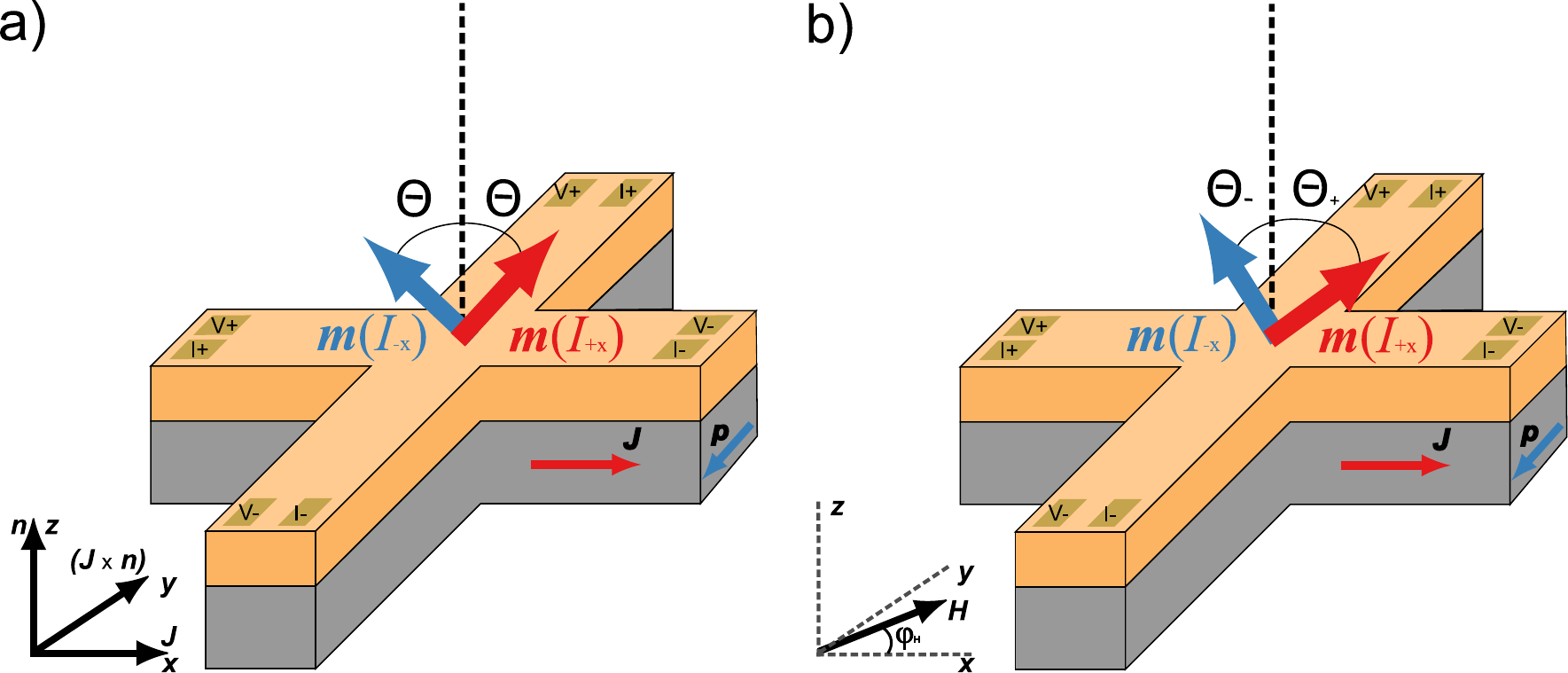}
\caption{Schematic illustration of the zero-offset magnetic field sensor. The ferromagnetic layer (orange, in experiment $\mathrm{Co_{60}Fe_{20}B_{60}}$) is situated on top of the heavy metal layer (gray, in experiment Ta) which acts as a spin-polarizer. Upon application of a charge current through the HM-layer $\boldsymbol{J}$ a spin-polarized current flows along $\boldsymbol{n}$ into the FM layer with a polarization $\boldsymbol{p} =  \left(\boldsymbol{J}\times\boldsymbol{n}\right)$. The behavior of the single domained OOP magnetization is illustrated first in a) with a vanishing external field, where the initial magnetization $\boldsymbol{m_0}$ is now tilted to $\boldsymbol{m}(I_{+x})$ for a positive charge current, and to $\boldsymbol{m}(I_{-x})$ for a negative charge current, respectively. The deviation angle $\Theta$ is equal for both current directions. The change induced by an external field applied in-plane is illustrated in b) where the presence of the external field disturbs the symmetry of SOT. The resulting angles $\Theta_+$ and $\Theta_-$ differ now in magnitude. }
\label{fig:concept}
\end{figure*}

The proposed sensor concept is based on a Hall cross with a heavy metal underlayer and a magnetic top layer with perpendicular magnetic anisotropy (PMA), see Fig. 1. Subject to an electric current in $\pm x$-direction, the spin Hall effect \cite{hirsch_spin_1999} in the heavy-metal layer leads to a spin current with a polarization in  $\mp y$-direction flowing into the magnetic layer. This leads to a spin-torque referred to as spin-orbit torque (SOT) \cite{manchon_current-induced_2019}.

For a vanishing external field and charge current, the equilibrium magnetization in the magnetic layer is perfectly aligned out-of-plane (OOP) due to the PMA. The response of the magnetic system to the spin-orbit torque is perfectly symmetric \cite{garello_symmetry_2013} when changing the sign of the electric current in the heavy metal layer, see Fig. 1(a). However, if an external field is applied, the equilibrium position of the magnetization is tilted and the response to the SOT becomes asymmetric, see Fig. 1(b).

The main idea of the proposed sensing principle is to perform differential measurements with alternating currents in order to exploit this symmetry and eliminate the offset for the zero-field case.

\section{\label{sec:level3}MODELING}
In order to verify the feasibility of the proposed concept, we perform a series of macrospin simulations where the spin-orbit-torque interactions with the magnetization are described by the Landau-Lifshitz-Gilbert-Equation (LLG)\cite{gilbert_phenomenological_2004, abert_micromagnetics_2019}

\begin{equation}
\centering
{{\partial}_t \boldsymbol{m}} = {-\gamma\boldsymbol{m}\times{\boldsymbol{H}}^{\mathrm{eff}}+\alpha{\boldsymbol{m}}\times{\partial}_t{\boldsymbol{m}}+\boldsymbol{T}_{\mathrm{damp}}+\boldsymbol{T}_{\mathrm{field}}},
\label{eq:LLG}
\end{equation}

where $\gamma$ is the gyromagnetic ratio, $\boldsymbol{H}^{\mathrm{eff}}$ the effective field of the FM-layer and $\alpha$ the Gilbert-damping-coefficient.
The LLG is extended by the two additional current induced torques~\cite{slonczewski_currents_2002, abert_fieldlike_2017, abert_micromagnetics_2019}

\begin{equation}
\label{eq:DT}
\boldsymbol{T}_{\mathrm{damp}}={\eta}_{\mathrm{damp}}\frac{j_e\gamma\hbar}{2e{\mu}_{0}tM_s} \boldsymbol{m} \times  (\boldsymbol{m} \times  \boldsymbol{p}),
\end{equation}

and 

\begin{equation}
\label{eq:FT}
\boldsymbol{T}_{\mathrm{field}}={\eta}_{\mathrm{field}}\frac{j_e\gamma\hbar}{2e{\mu}_{0}tM_s} \boldsymbol{m} \times  \boldsymbol{p},
\end{equation}
with $\hbar$ being the reduced Planck-constant, $e$ being the elementary charge, and $\mu_0$ being the vacuum permeability.
The strength of the two torques is given by  the magnitude of the charge current $j_e$ applied in the heavy metal layer, parameters of the magnetic system such as saturation magnetization $M_s$, thickness $t$, as well as the dimensionless SOT coefficients $\eta_\mathrm{damp}$ and $\eta_\mathrm{field}$. 

Equation~\eqref{eq:LLG} can then be written as

\begin{equation}            
\begin{aligned}
\centering
{\partial}_t \boldsymbol{m} = -\gamma \boldsymbol{m}\times\left(\boldsymbol{H^{\mathrm{eff}}} - H_{\mathrm{dl}} \boldsymbol{m} \times  \boldsymbol{p} - H_\mathrm{fl}\boldsymbol{p}
\right) + \alpha{\boldsymbol{m}}\times{\partial}_t{\boldsymbol{m}},
\end{aligned}
\label{eq:SOT}
\end{equation}

where

\begin{equation}
\label{eq:Hdl}
H_\mathrm{dl} = \dfrac{j_e\hbar}{2e\mu_0tM_s} \cdot \eta_\mathrm{dl},
\end{equation}

and

\begin{equation}
\label{eq:Hfl}
H_\mathrm{fl} = \dfrac{j_e\hbar}{2e\mu_0tM_s} \cdot \eta_\mathrm{fl}
\end{equation}
are the magnitudes of the current induced damping-like and field-like torques.

\subsection{\label{sec:level3B}First-order analytical approximation}
The magnetization dynamics of a perpendicularly magnetized system can be described by macrospin simulations where Eq.~\eqref{eq:SOT} is solved numerically, or by employing an analytical approach like Hayashi et al\cite{hayashi_quantitative_2014}. Here, we derive an even simpler first-order approximation based on the Stoner-Wolfarth formalism. 

Let us assume now that the ferromagnetic layer is homogeneously magnetized along the [001] direction, $m_0 = (0,0,1)^T$. When a homogeneous magnetic field is applied in the $xy$ plane, the effective field $H^{\mathrm{eff}}$ is given solely by the contribution from the anisotropy and Zeeman energies \cite{abert_micromagnetics_2019} as

\begin{equation}
\label{eq:Heff}
H^\mathrm{eff} = 
\begin{pmatrix} 
H_x  \\
H_y \\
H_k 
\end{pmatrix},
\end{equation}
where $H_x$, and $H_y$ are the components of the applied in-plane field, and $H_k$ is the effective anisotropy field.

In order to account for the spin-orbit torque, the effective field has to be complemented according to Eq.~\eqref{eq:SOT}. In the case of a spin-polarization in $y$-direction, the field-like torque only contributes via the $y-$component of the effective field

\begin{equation}
\label{eq:Heffy}
H^{\mathrm{eff}}_{y} = H_y - p_y\cdot H_\mathrm{fl},
\end{equation}

whereas the damping-like torque contributes to the $x$-component as

\begin{equation}
\label{eq:Heffx}
H^{\mathrm{eff}}_{x} = H_x - (\boldsymbol{m}\times \boldsymbol{p})_x\cdot  H_\mathrm{dl}.
\end{equation}

The magnitude of the external field applied along the sensitive direction and SOT fields is then given by
\begin{equation}
H_{\parallel} = \sqrt{(H^{\mathrm{eff}}_{x})^2 + (H^{\mathrm{eff}}_{y})^2}.
\end{equation}

According to the hard-axis approximation of the Stoner-Wohlfarth model, the parallel component of the magnetization is given as

\begin{equation}
m_{\parallel} = \dfrac{H_{\parallel}}{H_k},
\end{equation}
where $H_k$ is the anisotropy field.

The $z$ component of the magnetization can then be simply obtained from the unit magnetization constraint where
\begin{equation}
\label{eq:mz}
m_z = \pm \sqrt{ 1 - m_{\parallel}^2},
\end{equation}
with the sign depending on the initial sign of $m_z$.

Consider now, that we apply a DC current $I_0$ along $+x$-direction. For the sake of simplicity we refer to this current as $I_{+x}$. If the current is applied along $-x$-direction, $I_{-x} = -I_0$. As explained in Sec.~\ref{sec:level2} this leads to a change in magnetization, with the new magnetization state being $m_z(I_{+x})$, which is proportional to the anomalous Hall resistance

\begin{equation}
\label{eq:Rxy}
	R_{xy}(I_{+x}) = \Delta R_{AHE}m_z(I_{+x}),
\end{equation}
where $\Delta R_\mathrm{AHE}$ is the anomalous hall coefficient. Thus, one obtains the hall voltage $V_{xy}(I_{+x}) = \Delta R_{AHE}m_z(I_{+x})I_{+x}$.

The sensor signal can be obtained either as the difference of measured resistances
 
\begin{equation}
\label{eq:signal_R}
 S = R_{xy}(I_{+x}) - R_{xy}(I_{-x}),
\end{equation}

or as a sum of measured AHE voltages

\begin{equation}
\label{eq:signal_V}
S_{xy} = V_{xy}(I_{+x}) + V_{xy}(I_{-x}).
\end{equation}

The linearity of the sensor sensor signal $S$ can be derived from Eq.~\eqref{eq:mz} directly.
Let us now assume the most general case, where both SOT coefficients are included, and thus both $x,$ and $y$ external fields are applied to be aligned with the sensitive direction. We can then write

\begin{equation}
	\label{eq:Vxy_0}
	V_{xy}(I_{+x}) = \pm\Delta R_{AHE}I_{+x}\sqrt{1-\dfrac{(H_x \mp H_\mathrm{dl})^2 + (H_y + H_\mathrm{fl})^2}{H_k^2}}.
\end{equation}

After expanding Eq.\eqref{eq:Vxy_0} in a Taylor series at $H_x=0$ and $H_y=0$ and summing up the resulting voltages given in Appendix~Eq.~\ref{eq:TaylorAll} according to Eq.~\eqref{eq:signal_V}, one obtains the simplified analytical approximation of the sensor signal where

\begin{equation}
\label{eq:Sxy_Hxy}
S_{xy} = -2\Delta R_{AHE}I_{0}\left( H_\mathrm{dl}\dfrac{H_x}{{H_k}^2} \mp H_\mathrm{fl}\dfrac{H_y}{{H_k}^2}\right)  + \mathcal{O}[H_x]^2 + \mathcal{O}[H_y]^2.
\end{equation}

The field components $H_x$ and $H_y$ can be written as $H_x = H\cos{\phi_H}$ and $H_y = H\sin{\phi_H}$, respectively. Hence, the total sensor signal becomes

\begin{equation}
\label{eq:Sxy}
S_{xy} = -2\Delta R_{AHE}I_{0}\dfrac{H( H_\mathrm{dl}\cos{\phi_H} \mp H_\mathrm{fl}\sin{\phi_H})}{H_k^2} + \mathcal{O}[H_x]^2 + \mathcal{O}[H_y]^2.
\end{equation}

\subsection{\label{sec:level3C}Equivalency to higher harmonics voltage measurements}
For systems where very low signals are expected, or highly accurate measurements are desired, a higher harmonics voltage analysis can be performed\cite{, garello_symmetry_2013,hayashi_quantitative_2014,avci_current-induced_2017, schulz_effective_2017}. In this case, one applies an AC current instead of performing two subsequent measurements with opposite currents. The applied frequencies are in the Hz range.  

Consider again the perpendicularly magnetized system, where the magnetization is modulated due to SOTs. Upon application of the current $I(t) = I_0\sin(\omega t)$, the magnetization component $m_z$, and thus, $R_{xy}$ oscillates with the frequency $2\omega$. The hall voltage can be expressed as

\begin{equation}
	V_{xy}(t) = R_{xy}I(t).
\end{equation}

Substituting $R_{xy}$ with the  temporal evolution of Eq.~\eqref{eq:Rxy} one obtains 

\begin{equation}
	V_{xy}(t) = \Delta R_{\mathrm{AHE}}I(t)m_z(t),
\end{equation}
which can be further completed with the first order approximation for the $z-$component of the magnetization to

\begin{equation}
\label{eq:Vxyt0}
V_{xy}(t) =\pm \Delta R_{\mathrm{AHE}}I_0\sin{(\omega t)}\sqrt{1-\dfrac{(H_x \mp H_\mathrm{dl}(t))^2 + (H_y + H_\mathrm{fl}(t))^2}{H_k^2}}.
\end{equation}

As described in Appendix~\ref{ap:B} one can then perform a second order Fourier series expansion of Eq.~\eqref{eq:Vxyt0} with respect to $\omega t$, and sum up the terms accordingly to
 
\begin{equation}
\label{eq:Vxyt2}
\begin{aligned}
V_{xy}(t) = &\dfrac{I_0\Delta R_{AHE}\left(H_\mathrm{dl}H_{x}\mp H_\mathrm{fl}H_{y} \right)}{2{H_k}^2}\\
&\mp \dfrac{I_0\Delta R_{AHE} \left(3(H_\mathrm{dl}^2+H_\mathrm{fl}^2)-16H_k^2+4(H_{x}^2+H_{y}^2)\right)}{8H_k^2}\sin(\omega t)\\
&\pm \dfrac{I_0\Delta R_{AHE}\left(H_\mathrm{dl}H_{x}\mp H_\mathrm{fl}H_{y} \right)}{2{H_k}^2}\cos(2\omega t).
\end{aligned}
\end{equation}

Based on the common higher-harmonics voltage analysis where the time dependent AHE voltage is given as

\begin{equation}
V_{xy}(t) = V_0 + V_{\omega}\sin{(\omega t)} + V_{2\omega}\cos{(2\omega t)},
\end{equation}
one can express the second-harmonic voltage as
\begin{equation}
	V_{2\omega} = \pm \dfrac{I_0\Delta R_{AHE}\left(H_\mathrm{dl}H_{x}\mp H_\mathrm{fl}H_{y} \right)}{2{H_k}^2},
\end{equation}

which differs from our sensor signal approximation from Eq.~\eqref{eq:Sxy} only by a constant factor.

Thus, one can measure the sensor signal either by the sum of AHE voltages for positive and negative currents, by applying an AC current and measuring the DC contribution, or by measuring the field dependence of the second harmonics voltage using higher-harmonics voltage analysis (HHVA), e.g. with a lock-in amplifier.

\subsection{\label{sec:level3D}Micromagnetic Simulations}
In order to check the accuracy of the analytical model, we additionally perform fully dynamical LLG simulations. For the sake of simplicity, we use a single-cell approach for the simulation of the magnetic layer using the micromagnetic simulation software \texttt{magnum.np}\cite{bruckner_magnum.np_2023}.
For the effective field, we consider the effective anisotropy field and the external field only.
The effect of the demagnetization field is considered in terms of a contribution to the anisotropy field.
As material parameters, we have chosen the saturation magnetization $\mu_0 M_s=\SI{1.2}{T}$ and Gilbert damping constant $\alpha = 1$, as suitable material parameters for a Ta-CoFeB system~\cite{manchon_current-induced_2019}. Since the thickness of the magnetic layer is crucial for the strength of current induced torques, we assume everywhere a thickness of $t=\SI{1}{nm}$. To obtain the voltage from the magnetization component $m_z$ we consider a cross-section $A = L\times t = \SI{10}{\mu m} \times \SI{1}{nm}$ to calculate the applied current from the magnitude of the applied current density $j_e$ and we assume $\Delta R_{\mathrm{AHE}} = \SI{1}{\Omega}$. 

\section{\label{sec:level4}EXPERIMENTAL METHODS}
Our numerical investigations are accompanied by experimental validations, where a SOT stack consisting of an HM layer, Ta, and an FM Layer, CoFeB is used. Due to the equivalence showed in Sec.~\ref{sec:level3C}, we are measuring the second harmonics voltage with a Lock-in technique, and investigate the field-dependency of the sensor signal.

\subsection{Sample Fabrication}

The SOT structure was grown on an 8'' silicon wafer using a  Singulus Rotaris tool.  Before depositing the film stack, the wafers got an aluminum metallization layer to contact the SOT structure from the bottom. The connection of the SOT structure with the aluminum metal layer through an insulation $\mathrm{SiO_2}$ layer is realized with tungsten vias. In order to have a smooth surface before depositing the SOT film stack, a chemical mechanical polish was carried out.
The film stack consists of $\mathrm{Ta}[\SI{6}{nm}]/\mathrm{C_{60}Fe_{20}B_{20}}[\SI{1}{nm}]/\mathrm{MgO}[\SI{1.5}{nm}]/\mathrm{Ta}[\SI{5}{nm}]$. It was grown without vacuum brake by physical vapor deposition (PVD) with a base pressure $< 5\times 10^{-9}$ torr. The used sputter gas was Argon for all layers. Metal deposition was done in DC sputtering mode, and MgO deposition was performed in RF mode. 
After deposition, the wafers were annealed for $\SI{2}{h}$ at a temperature of $\SI{280}{^\circ C}$ in a vacuum. The patterning of the SOT structure can be done by reactive ion etching (RIE) with chlorine or by ion beam etching (IBE) whereas the latter needs an additional hard mask. The Hall bars used in this study were patterned from the films using conventional optical lithography and Ar ion etching. To prevent the SOT structure from corrosion a passivation layer was deposited. Note that a further encapsulation by aluminum-filled vias surrounding and metallization layers on top of the SOT structure leads to better heat dissipation. Finally, the pads were released by opening the passivation layer. Standard separation techniques were used (mechanical dicing). 

\subsection{Second Harmonics Measurements}
The harmonic voltages are measured using a standard 2nd harmonic lock-in amplifier detection-based technique, where we utilize a vector magnet to apply the magnetic field in different directions. We use standard wire bonding to connect the Hall bar device with the sample holder. The Hall bar is connected to 50 $\Omega$ resistance in series to measure the input sinusoidal current during the measurement. The schematic is shown in Fig.\ref{fig:concept}a). Prior to the harmonic measurements, the Hall bar device is pre-saturated along the $\pm z$-direction. We apply a sinusoidal voltage with constant amplitude ($V_{in}$) using a lock-in amplifier (Model number: HF2LI by Zurich instrument) to the Hall bar device with a reference frequency $~\SI{13.7}{Hz}$. Note that we will be giving current densities as input currents to obtain a better analogy to the numerical studies. Other two lock-in amplifiers (model numbers: 7265 and 7225) were used to measure the in-phase first and out-of-phase second harmonic voltage simultaneously.

\section{\label{sec:level5}Results}
\subsection{\label{sec:level5A}Role of SOT coefficients}
We begin our numerical investigation with a thorough investigation of the role of the SOT coefficients on the equilibrium states, as this will define the sensitive direction of our magnetic field sensor.
By using a single-cell approach as explained in Sec.~\ref{sec:level3D}, we vary the field-like and damping-like coefficients $\eta_{\mathrm{damp}}$ and $\eta_{\mathrm{field}}$ which will alter the strength of the torques exerted on the magnetization. For this purpose we are using an effective anisotropy field $\mu_0H_k = \SI{200}{mT}$, and apply an SOT current of magnitude $j_e = \SI{10}{MAcm^{-2}}$.

\begin{figure*}[]
	\centering
	\includegraphics[width=0.9\textwidth]{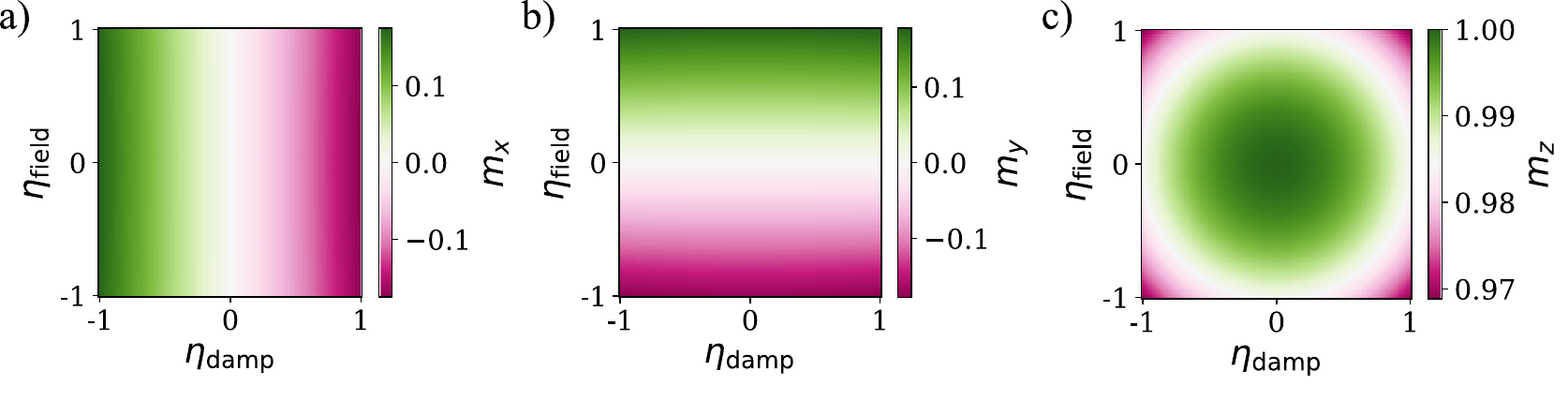}
	\caption{Dependence of the SOT coefficients on the equilibrium magnetization when an SOT current is applied to macrospin is shown where from a) we see that the damping like torque modulates the $x$-component. From b) is obvious that the field-like coefficient modulates the y-component of the magnetization, whereas from c) we learn that both torques contribute symmetrically to the modulation of the z-component of the magnetization.}
 	\label{fig:etas}
\end{figure*}

In Fig.~\ref{fig:etas} one can see the influence of the SOT coefficients on the equilibrium magnetization state. As expected, and briefly mentioned during the derivation of the first order analytical approximation, the variation of $\eta_{\mathrm{dl}}$ only changes the $x-$component of the magnetization, see Fig.~\ref{fig:etas}a), whereas the variation of $\eta_{\mathrm{fl}}$ modulates $m_y$, as it is illustrated in Fig.~\ref{fig:etas}b). Both torques contribute to the inclination of the magnetization in a symmetrical way, as shown in Fig.~\ref{fig:etas}c). In principle, the choice of the HM layer will lead to a specific pair of $\eta_{\mathrm{damp}}$ and $\eta_{\mathrm{field}}$\cite{manchon_current-induced_2019}, which in return will then modulate the sensitive direction of the sensor. The strength of the SOT coefficients can then be determined by measuring the first, and second harmonic voltages, and applying the Hayashi method for perpendicularly magnetized systems~\cite{hayashi_quantitative_2014}. If the planar hall-effect is neglectable, than the expression for $V_\omega$ and $V_{2\omega}$ given in Eq.~\eqref{eq:Vxyt2} can also be used. Alternative techniques such as Spin-Torque Ferromagnetic Resonance (ST-FMR)\cite{manchon_current-induced_2019} exist that lead to the precise determination of the damping-like coefficient.

We have chosen $\eta_{\mathrm{damp}}=-0.4$ and $\eta_{\mathrm{field}}=0.5$ for all the future investigations, such that the sensitive direction is slightly away from the [110] axis. The sensitive axis can be calculated also simply from the ratio of the two current induced torques given in Eqs.~\eqref{eq:Hdl} and \eqref{eq:Hfl} where $\tan{\phi_H} = H_{\mathrm{dl}}/H_{\mathrm{fl}}$. Note that the signs of the SOT coefficients depend on their definition in the LLG. In our case a positive field-like coefficient means, that the torque acts parallel to the Oersted field that is generated by the current flow in the SOT stack\cite{manchon_current-induced_2019}.

\subsection{\label{sec:level5B}Operation as a linear magnetic field sensor}
\begin{figure*}[]
	\centering
	\includegraphics[width=0.9\textwidth]{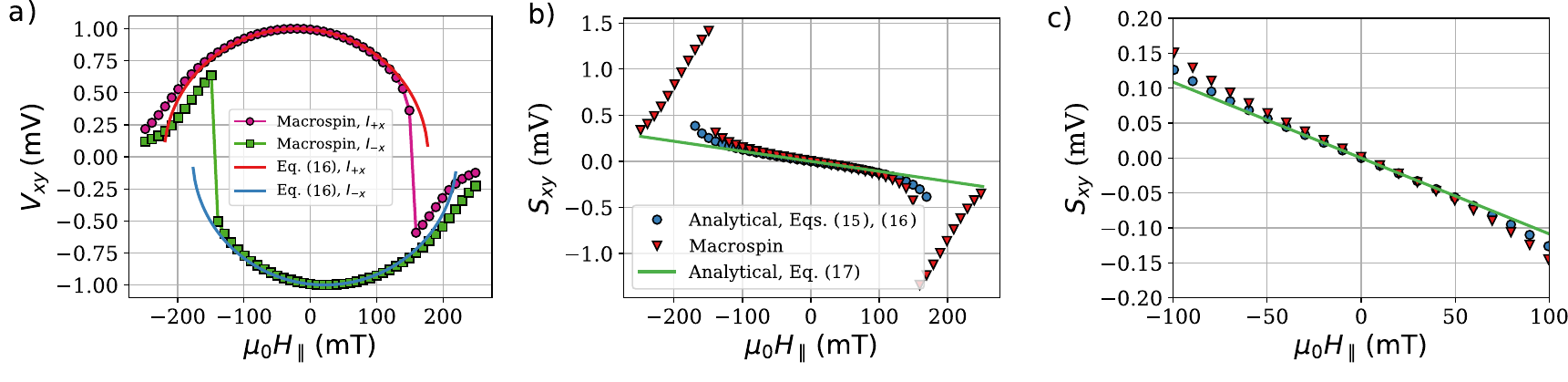}
	\caption{The simulated (pink, green) and calculated (red, blue) AHE voltages using simple approximation from Eq.\eqref{eq:mz} are depicted in a) for both current directions starting from $\boldsymbol{m}_0 = (0, 0, 1)^{\mathrm{T}}$. The calculated sensor signal is then plotted in b). A very good agreement between the analytical solutions, as well as the simulations can be observed in the limit of lower fields, which is highlighted in c), where we zoom into relevant range.}
 	\label{fig:sensor_signal}
\end{figure*}

Using our micromagnetic single cell simulations we apply the charge current subsequently in opposite directions, where $j_e = \SI{10}{MAcm^{-2}}$, and $\mu_0H_k = \SI{199}{mT}$. Now, we apply a magnetic field along the obtained sensitive axis $\boldsymbol{H}_\parallel$. Figure~\ref{fig:sensor_signal}a) shows the evolution of $V_{xy}$ as the response of the magnetization to the applied field. As in the simplified analytical model (Eq.~\eqref{eq:Vxy_0}, lines), the macrospin simulations (filled symbols) yield a parabolic dependence of $V_{xy}$ from the measured field $H_\parallel$. For opposite charge currents, the magnetization is modulated symmetrically around the vanishing magnetic field.  If the applied magnetic field exceeds a specific threshold, then the magnetization switches sign.

Making use of the proposed differential measurement, we obtain the sensor signal shown in Fig.~\ref{fig:sensor_signal}b). The sensor signal is obtained as the sum of Hall voltages. These signals are obtained from macrospin simulations (red), from Eq.~\eqref{eq:signal_V} (blue), and from Eq.~\eqref{eq:Sxy} (green line). A high linear range is observed, which is slightly smaller than the anisotropy field. Figure~\ref{fig:sensor_signal}c) illustrates the zoom of gray rectangle from b), which shows a good agreement between all three curves in the limit of small fields. The sensitivity of the sensing device, given by the slope of the sensor signal, starts to deviate for the analytical solutions significantly for higher fields, where we are approaching the anisotropy field. It is worth reminding that the small field limit was an approximation we have done in Sec.~\ref{sec:level3B}. Investigation of the transfer curves clearly yields offset-free sensor signals for all theoretical models. 

\begin{figure}[]
	\centering
	\includegraphics[width=\columnwidth]{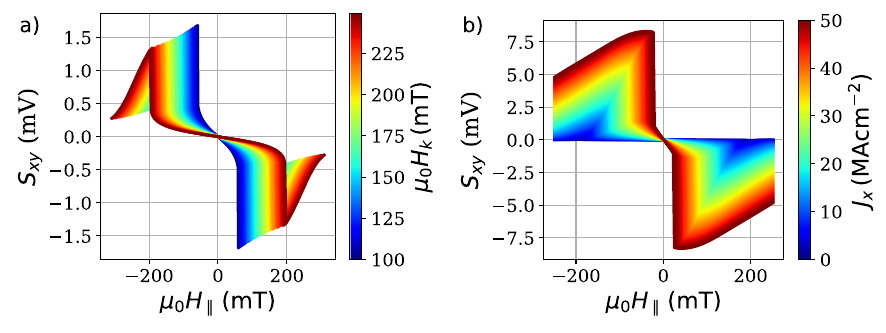}
	\caption{Dependence of the sensor signal on the strength of the perpendicular magnetic anisotropy is shown in a), where the linear range increases with $\mu_0H_k$, but to the cost of the sensitivity. The applied charge current density is depicted in b), where one can see that a higher current increases the sensitivity of the sensor signal, but in exchange it reduces the measurable linear range.}
	\label{fig:varje_Hk}
\end{figure}

To better understand the sensing performance, we are investigating the role of the effective anisotropy $\mu_0H_k$, as well as of the applied SOT current $J_x$. First, we vary $\mu_0H_k$, while keeping the applied current constant at $je = \SI{10}{MAcm^{-2}}$. The numerically obtained sensor signals are illustrated in Fig.~\ref{fig:varje_Hk}. We observe that it is possible to modulate both sensitivity and linear range of the sensor by the variation of the anisotropy field. The strength of the effective anisotropy field can be tuned via the thickness of the FM Layer (if the PMA originates as an interface effect). In materials like Ta-CoFeB or Pt-Co, which are the materials of choice for investigation of SOT, the anisotropy field at room temperature can vary between $\SI{50}{mT}$ to $\SI{300}{mT}$ \cite{manchon_current-induced_2019}. Based on desired application it is possible in principle to tune the linear range by either a change of materials, or adjusting the thickness of the FM layer.  

A second measure to tune the sensitivity and linear range of the sensor is the magnitude of the electric current applied to the heavy metal layer, see Fig.~\ref{fig:varje_Hk}b). At higher currents the magnetization switches much faster into the in-plane state due to the stronger current induced torques, limiting the linear range to be used for magnetic field detection. Higher modulations can be obtained using currents below the switching threshold, leading to the increase of the sensitivity. Note that a similar effect could be obtained by changing the material acting as HM, in order to scale the exerted torques via the SOT coefficients, or alternative spin polarizing materials can be chosen that use different physics, like the orbital orbit torque effect\cite{ding_harnessing_2020}. However, the dependence of the sensitivity on the applied current allows us to obtain a magnetic field sensor with tunable sensitivity, and linear range.
Independent of the chosen parameters the sensing signal remains always offset free.
  
\subsection{\label{sec:level5C}Sensor performance}
\begin{figure}[b]
	\centering
	\includegraphics[width=\columnwidth]{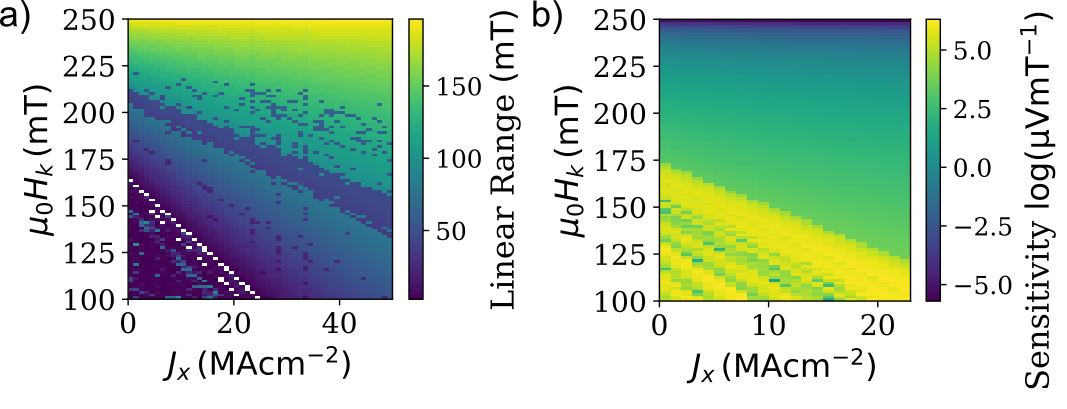}
	\caption{Magnetic field sensing properties as a function of applied SOT current $J_x$ and effective magnetic anisotropy$\mu_0H_k$, where the measurable linear range is shown in a), and the sensitivity is shown in b).}
	\label{fig:performance}
\end{figure}

Magnetic field sensors can be best compared on their properties like linear range, offset, or sensitivity. Hence, we are interested in the dependence of the linear range, and sensitivity on the applied current density $J_x$ and anisotropy field $\mu_0H_k$. The latter is a measure for the used material.
Figure \ref{fig:performance}a) illustrates the dependence of the chosen effective anisotropy and applied current on the obtained linear measurement range of magnetic sensor. From Fig.~\ref{fig:performance} it can be seen that the linear range is very well tunable with the change of the $\mu_0H_k$, and $J_x$.
Considering the sensitivity of the sensor signal depicted in Fig.~\ref{fig:performance}b), we distinguish between two phases. Above a certain given threshold of anisotropy and current the sensitivity drops significantly, and it cannot be tuned with the current anymore.

Another important measure for the performance of the sensor is its robustness against field components perpendicular to the measured field direction. In general, an arbitrary magnetic field will have an in-plane component that can be projected along the sensitive direction $H_\parallel$, and a normal component $H_\perp$, which might change the equilibrium state at vanishing fields as well, leading to poorer sensor performance.

In order to investigate the influence of perpendicular field components, we vary the strength of the normal component $H_\perp$, as well as the applied angle $\phi$. We first calculate the sensitive direction, normalize it, and rotate the normalized vector $90^\circ$ around the $z$-axis. Thus, when $\phi = 0$, and $m_\parallel = (1, 1, 0)$, $H_\perp$ is applied along $(-1, 1, 0)$.

Figure~\ref{fig:performance}a) reveals that a perturbation field does not influence the sensor offset, independent of amplitude and angle. Note that if $H_\perp > H_k$ then the magnetization will point along $H_\perp$ compromising the general sensing concept. A weak dependence of the sensitivity on the amplitude of $H_\perp$, and $\phi$ can be seen in Fig.~\ref{fig:performance}. When the normal field is applied, the sensitivity decreases as the equilibrium state will be slightly modulated if positive or negative currents are applied. However, there exists again a perturbation field threshold, $\mu_{0}H_\perp < \SI{10}{mT}$, where the sensitivity of the sensor signal remains rather constant and independent of the angle of the magnetization.

\begin{figure}
	\centering
	\includegraphics[width=\columnwidth]{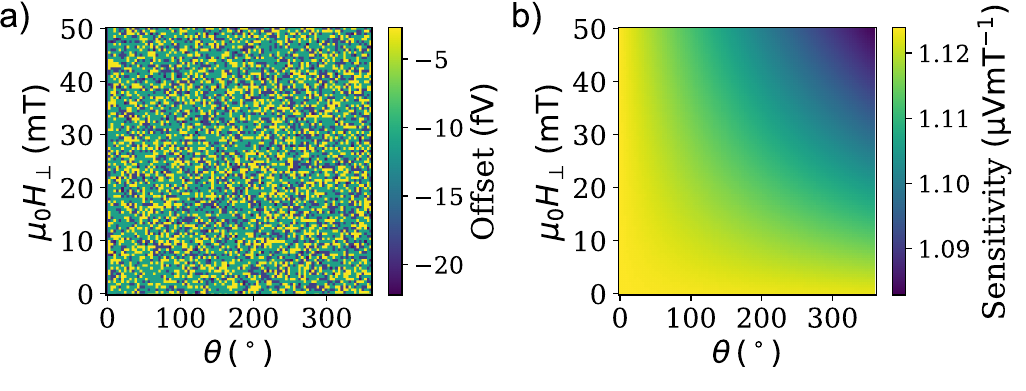}
	\caption{Influence of the external bias fields on the offset of the sensor signal (a) and on the sensor sensitivity (b). While the sensor offset remains in principle always offset free, the sensitivity changes slightly, yet insignificantly, with the amplitude of the perpendicular field $H_\perp$.}
 	\label{fig:performanceDist}
\end{figure}

\subsection{\label{sec:level5D}Higher Harmonics Voltage Measurements}
\begin{figure*}
	\centering
	\includegraphics[width=\textwidth]{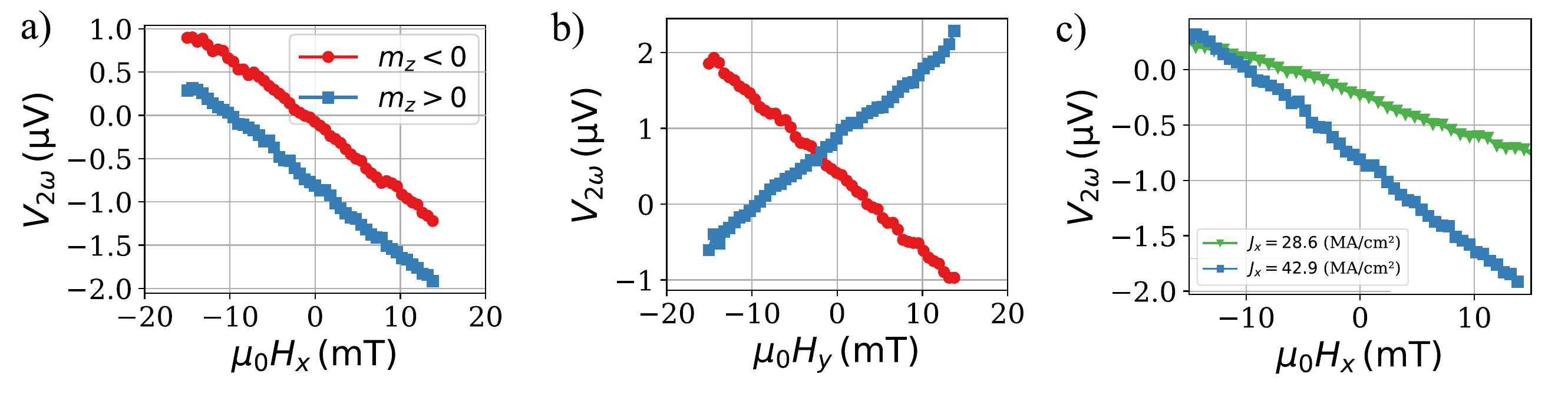}
	\caption{Experimental results obtained from the second harmonics measurements, where first the field parallel to current direction ($H_x$) is swept starting from both magnetization states, are shown in a). The transversal field is swept in b). The tunable sensitivity can be observed in c) where we sweep the longitudinal field for the charge current densities $J_x = \SI{28.6}{MAcm^{-2}}$ (green, triangles) and $J_x = \SI{42.9}{MAcm^{-2}}$ (blue, squares).}
	\label{fig:experiment}
\end{figure*}

To demonstrate the developed sensor concept, we have fabricated a Hall-cross structure, as explained in Sec.~\ref{sec:level4}. Using the anomalous Hall effect one can measure the change in $m_z$. Since our sensor is based on a differential measurement in order to eliminate the sensing offset, we employ second harmonics measurements, as discussed in Sec.~\ref{sec:level4}B.  Figure~\ref{fig:concept}a) illustrates the considered Hall structure. The measured $V_{2\omega}$ curves are plotted as a function of the applied magnetic field in Fig.\ref{fig:experiment}. While the dependence of the $V_{2\omega}$ on $H_x$ is given in Fig.~\ref{fig:experiment}a), the dependence of $H_y$ is illustrated in Fig.~\ref{fig:experiment}b). Red (blue) curves show the signal if the sample was pre-saturated along $+z$ ($-z$) directions using an OOP magnetic field. The charge current density for SOT sensing was $J_x = \SI{28.6}{MAcm^{-2}}$.

As expected from our simple analytical derivation, as well as from the derivations of Hayashi\cite{hayashi_quantitative_2014}, the $2\omega$ signal should change signs when the magnetic field is applied along the current direction if the initial magnetization state changes signs. If the magnetic field is applied perpendicular to the current direction, i.e. $H_y$, then the slope of the measured signal changes, as depicted in Fig.~\ref{fig:experiment}.

If one decreases the applied SOT current then the sensitivity of the sensor signal decreases as theoretically described above, and experimentally demonstrated and depicted in Fig.~\ref{fig:experiment}c).

Overall, all signals are linearly depending on the applied magnetic field, thus allowing us the experimental validation of sensor principle using SOT. However, a significant offset is observed if the magnetization was pre-saturated along $-z$ direction. While in experimental works where the HHVA is performed one usually disregards this offset attributing them to electronics, it limits the sensing performance of our single-device magnetic field sensor. 
Part of the offset can originate from the electronics in the measurement setup. However, also the sample can contribute to the offset due to the anomalous Nernst effect, which gives a contribution to the AHE and can be quantified. With this correction (that we are not plotting here to show the real measurement data) only a very minor offset remains, which we attribute to the electronics of the system.

Consider now that current flow in $x-$direction causes an increase in temperature due to Joule heating. Assuming that the change in temperature is small enough to not alter any material parameters, it is expected that we will have temperature gradients ($\nabla T$). 

When the magnetization $\boldsymbol{m}$ and $\nabla T$ are normal to each other, then the anomalous Nernst effect (ANE) gives an additional contribution due to the generated electrical field\cite{mizuguchi_energy-harvesting_2019}
\begin{equation}
	\boldsymbol{E}_{\mathrm{ANE}} = Q_s\mu_0M_s\left(\boldsymbol{m}\times\nabla T\right),
\end{equation}

where $Q_s$ is the ANE coefficient.

Let us now consider that the gradient of temperature has only a $z$ component, and we have an arbitrary magnetization state. Thus, the ANE generates an electric field given by

\begin{equation}
\boldsymbol{E}_{\mathrm{ANE}}(I_{+x})=Q_s\mu_0M_s
\begin{pmatrix} 
m_x  \\
m_y \\
m_z 
\end{pmatrix}
\times
\begin{pmatrix} 
0 \\ 
0 \\
\partial_z T 
\end{pmatrix}
=
\begin{pmatrix} 
m_y\partial_z T \\ 
-m_x\partial_z T \\
0
\end{pmatrix}.
\label{eq:E_ANE_vec}
\end{equation}
The measured hall voltage then becomes 
\begin{equation}
V_{xy}(I_{+x}) = V_\mathrm{AHE}(I_{+x}) + V_\mathrm{ANE}(I_{+x}),
\end{equation}
where 

\begin{equation}
	V_\mathrm{ANE}(I_{+x}) = wE_y(I_{+x}) = -wQ_s\mu_0M_sm_x\partial_z T
\end{equation}

For the opposite current the magnetization changes signs for the $x,$ and $y$ components leading to
\begin{equation}
V_\mathrm{ANE}(I_{-x}) = wE_y(I_{-x}) = wQ_s\mu_0M_sm_x\partial_z T.
\end{equation}

Summing up the measured hall voltages for the sensing signal according to Eq.\ref{eq:signal_V} yields that contributions from the ANE due to $\partial_z T$ are vanishing. Thus, we conclude that the z-component of the temperature gradient, that originates in the dissipation of the Joule heating, does not lead to any offset.

Let us now consider that the one obtains a gradient in temperature where
\begin{equation}
	\nabla T
	=
	\begin{pmatrix} 
	\partial_x T \\ 
	0 \\
	0
	\end{pmatrix},
\end{equation}
which can originate in the imperfections in current flow, and small symmetry distortion, e.g. slightly non-symmetrical arrangements of vias due to lithography tolerances.

The ANE contribution to the hall voltage then becomes
\begin{equation}
V_{ANE}(I_{+x}) = wE_y(I_{+x}) = wQ_s\mu_0M_sm_z\partial_x T.
\end{equation}

Changing the sign of the magnetization leads to the same term
\begin{equation}
V_{ANE}(I_{-x}) = wE_y(I_{-x}) = wQ_s\mu_0M_sm_z\partial_x T,
\end{equation}
which leads to the offset $2wQ_s\mu_0M_sm_z\partial_x T$ when the measured voltages are added. These considerations indicate, that the nonvanishing offset obtained in the experiments can probably be attributed to ANE. 
However not due to the z component of the gradient caused by the cooling of the sample towards the heat sink, as we have initially expected, but by the broken symmetry of the hall-bar due to fabrication defects, tolerances, or due to misalignments.

\section{\label{sec:level6}CONCLUSION}
In this work, we propose a novel concept for magnetic field sensing based on spin-orbit-torques. In essence, the proposed concept is the equivalent to the experiments performed to extract the SOT induced torques, as in the well established Hayashi method\cite{hayashi_quantitative_2014}.

In this sensing concept, the SOT leads to a symmetric inclination of the initial out-of-plane magnetization for opposite current directions.
We investigate the roles of the SOT-coefficients $\eta_\mathrm{damp}$ and $\eta_\mathrm{field}$, which are a measure of the damping-like and field-like torques respectively.

Our numerical investigations reveal that the sensor signal evolves linearly with the applied magnetic field in the working regime. This sensing principle shows a large linear range (hundreds of mT), whereas the offset vanishes in the absence of an external field. The sensitivity, as well as the linear range, can be tuned by the variation of the amplitude of the charge current, and anisotropy fields, respectively.

Perturbation fields applied in the normal plane to the sensitive direction do not affect the elimination of the offset, however, they slightly change the sensitivity of the sensor, and thus, lead to a poorer sensor performance.

The performed second harmonics measurements of the AHE allowed us to deliver a proof-of concept that the proposed sensing method is applicable, and second harmonics analysis can be performed to obtain a linear sensing signal. Furthermore, the increase in the applied current leads to a higher sensitivity allowing for the realization of a tunable modular sensor. However, the experimental validations show that the AHE is not perfectly suitable to measure the change in the magnetization, as the anomalous Nernst effect contribution leads to an offset. This additional offset can be eliminated if the magnetization is read through a TMR stack\cite{suess_device_2021}. 

\appendix
\section{\label{ap:A}Taylor Expansion of Sensor Signal}
One can expand Eq.~\eqref{eq:Vxy_0} in a Taylor series first at $H_x = 0$, which yields

\begin{equation}
\label{eq:Vxy_taylor_+x}
\begin{aligned}
&V_{xy}(I_{+x})\lvert{_{H_x = 0}} =  \Delta R_{AHE}I_{+x}\sqrt{-\dfrac{-H_\mathrm{dl}^2 - H_\mathrm{fl}^2 - 2H_\mathrm{fl}H_y+H_k^2-H_y^2}{H_k^2}}\\
& + \Delta R_{AHE}I_{+x}H_\mathrm{dl}H_x\dfrac{\sqrt{\dfrac{-H_\mathrm{dl}^2 - H_\mathrm{fl}^2 - 2H_\mathrm{fl}H_y+H_k^2-H_y^2}{H_k^2}}}{{H_k}^2\sqrt{1+\dfrac{-H_\mathrm{dl}^2 - (H_\mathrm{fl}+H_y)^2}{H_k^2}}} + \mathcal{O}[H_x]^2.
\end{aligned}
\end{equation}

For the opposite current $I_{-x}$, we obtain
\begin{equation}
\label{eq:Vxy_taylor_-x}
\begin{aligned}
&V_{xy}(I_{-x})\lvert{_{H_x = 0}} =  \Delta R_{AHE}I_{-x}\sqrt{-\dfrac{-H_\mathrm{dl}^2 - H_\mathrm{fl}^2 + 2H_\mathrm{fl}H_y+H_k^2-H_y^2}{H_k^2}}\\
& - \Delta R_{AHE}I_{-x}H_\mathrm{dl}H_x\dfrac{\sqrt{\dfrac{-H_\mathrm{dl}^2 - H_\mathrm{fl}^2 + 2H_\mathrm{fl}H_y+H_k^2-H_y^2}{H_k^2}}}{-H_\mathrm{dl}^2 - H_\mathrm{fl}^2 + 2H_\mathrm{fl}H_y+H_k^2-H_y^2} + \mathcal{O}[H_x]^2.
\end{aligned}
\end{equation}

If one evaluates Eq.~\ref{eq:Vxy_0} at $H_y=0$, this yields
\begin{equation}
\label{eq:VxyHy_taylor_+x}
\begin{aligned}
&V_{xy}(I_{+x})\lvert{_{H_y = 0}} =  \Delta R_{AHE}I_{+x}\sqrt{-\dfrac{-H_\mathrm{dl}^2 - H_\mathrm{fl}^2 + 2H_\mathrm{dl}H_x+H_k^2-H_x^2}{H_k^2}}\\
& - \Delta R_{AHE}I_{+x}H_\mathrm{fl}H_y\dfrac{\sqrt{\dfrac{-H_\mathrm{dl}^2 - H_\mathrm{fl}^2 + 2H_\mathrm{dl}H_x+H_k^2-H_x^2}{H_k^2}}}{-H_\mathrm{dl}^2 - H_\mathrm{fl}^2 + 2H_\mathrm{dl}H_x+H_k^2-H_x^2} + \mathcal{O}[H_y]^2.
\end{aligned}
\end{equation}
for positive current $I_{+x}$ and
\begin{equation}
\label{eq:VxyHy_taylor_-x}
\begin{aligned}
&V_{xy}(I_{-x})\lvert{_{H_y = 0}} =  \Delta R_{AHE}I_{-x}\sqrt{-\dfrac{-H_\mathrm{dl}^2 - H_\mathrm{fl}^2 - 2H_\mathrm{fl}H_y+H_k^2-H_y^2}{H_k^2}}\\
& + \Delta R_{AHE}I_{-x}H_\mathrm{fl}H_y\dfrac{\sqrt{\dfrac{-H_\mathrm{dl}^2 - H_\mathrm{fl}^2 - 2H_\mathrm{fl}H_y+H_k^2-H_y^2}{H_k^2}}}{{H_k}^2\sqrt{1+\dfrac{-H_\mathrm{fl}^2 - (H_\mathrm{dl}+H_x)^2}{H_k^2}}} + \mathcal{O}[H_y]^2,
\end{aligned}
\end{equation}
for negative current $I_{-x}$.

We can then simplify $\sqrt{-\dfrac{-H_\mathrm{dl}^2 - H_\mathrm{fl}^2 + 2H_\mathrm{dl}H_x+H_k^2-H_x^2}{H_k^2}}\approx 1$, since in the first order approximation one has to consider $-H_\mathrm{dl}^2 - H_\mathrm{fl}^2 + 2H_\mathrm{dl}H_x-H_x^2\ll H_k^2$, which leads to even further simplifications, where
\begin{equation}
\label{eq:TaylorAll}
\begin{aligned}
& V_{xy}(I_{+x})\lvert{_{H_x = 0}} = \Delta R_{AHE}I_{0}
+ \Delta R_{AHE}I_{0}\dfrac{H_\mathrm{dl}Hx}{H_k^2} + \mathcal{O}[H_x]^2,\\
& V_{xy}(I_{-x})\lvert{_{H_x = 0}} = -\Delta R_{AHE}I_{0}
+ \Delta R_{AHE}I_{0}\dfrac{H_\mathrm{dl}Hx}{H_k^2} + \mathcal{O}[H_x]^2\\
& V_{xy}(I_{+x})\lvert{_{H_y = 0}} = \Delta R_{AHE}I_{0}
- \Delta R_{AHE}I_{0}\dfrac{H_\mathrm{fl}Hy}{H_k^2} + \mathcal{O}[H_y]^2,\\
& V_{xy}(I_{-x})\lvert{_{H_y = 0}} = -\Delta R_{AHE}I_{0}
- \Delta R_{AHE}I_{0}\dfrac{H_\mathrm{fl}Hy}{H_k^2} + \mathcal{O}[H_y]^2.
\end{aligned}
\end{equation}

Summing up all the Taylor expansions one can then obtain the the sensor signal according to Eq.~\eqref{eq:signal_V}.

\section{\label{ap:B}Fourier Expansion of AC Sensor Signal}
The measured time-dependent voltage output $V_{xy}(t)$ upon application of an AC SOT current can be divided into four cases. If one starts with an initial magnetization pointing along $[001]$, i.e. $m_z>0$, and the the magnetic field being applied along $x-$direction, the AHE voltage is given by
\begin{equation}
\label{eq:B1}
	V_{xy}(t) = \Delta R_{AHE}I_0\sin{(\omega t)}\cdot \sqrt{1 - \dfrac{\left(H_x - H_\mathrm{dl}(t)\right)^2}{H_k^2}},
\end{equation}
where the current induced damping-like torque is now time-dependent and becomes
 \begin{equation}
 \label{eq:B2}
 	H_\mathrm{dl}(t) = H_\mathrm{dl}\sin{(\omega t)},
 \end{equation}
with $\omega$ being the frequency of the applied AC current.
Thus one can write the Eq.~\eqref{eq:B1} as
\begin{equation}
\label{eq:B3}
V_{xy}(t) = \Delta R_{AHE}I_0\sin{(\omega t)}\cdot \sqrt{1 - \dfrac{\left(H_x - H_\mathrm{dl}\sin{(\omega t)}\right)^2}{H_k^2}}.
\end{equation}
Using the approximation $\sqrt{1 -\frac{a}{b^2}} \approx \left(1-\frac{a}{2b^2}\right)$, for $(a \ll b^2)$, one can further simplify Eq.~\eqref{eq:B3} to
\begin{equation}
\label{eq:B4}
V_{xy}(t) = \Delta R_{AHE}I_0\sin{(\omega t)}\cdot {\left(1 - \dfrac{\left(H_x -H_\mathrm{dl}\sin{(\omega t)}\right)^2}{2H_k^2}\right)}.
\end{equation}
The second order Fourier expansion of Eq.~\eqref{eq:B4} at $\omega t$ can then be written as

\begin{equation}
	\label{eq:B5}
	\begin{aligned}
	V_{xy}(t) = &\dfrac{I_0\Delta R_{AHE}H_\mathrm{dl}H_{x}}{2{H_k}^2}\\
				&-\dfrac{I_0\Delta R_{AHE}\left(3H_\mathrm{dl}^2-8H_k^2+4H_{x}^2\right)}{8H_k^2}\sin(\omega t)\\
				&- \dfrac{I_0\Delta R_{AHE}H_\mathrm{dl}H_{x}}{2{H_k}^2}\cos(2\omega t).
	\end{aligned}
\end{equation}

Considering that one starts with the opposite magnetization state ($[00\bar{1}]$, $m_z$<0), the the Fourier expansion becomes
\begin{equation}
\label{eq:B6}
\begin{aligned}
V_{xy}(t) = &\dfrac{I_0\Delta R_{AHE}H_\mathrm{dl}H_{x}}{2{H_k}^2}\\
&+\dfrac{I_0\Delta R_{AHE}\left(3H_\mathrm{dl}^2-8H_k^2+4H_{x}^2\right)}{8H_k^2}\sin(\omega t)\\
&- \dfrac{I_0\Delta R_{AHE}H_\mathrm{dl}H_{x}}{2{H_k}^2}\cos(2\omega t).
\end{aligned}
\end{equation}

Thus, as a general case on can then write 
\begin{equation}
\label{eq:B7}
\begin{aligned}
V_{xy}(t) = &\dfrac{I_0\Delta R_{AHE}H_\mathrm{dl}H_{x}}{2{H_k}^2}\\
&\mp\dfrac{I_0\Delta R_{AHE}\left(3H_\mathrm{dl}^2-8H_k^2+4H_{x}^2\right)}{8H_k^2}\sin(\omega t)\\
&- \dfrac{I_0\Delta R_{AHE}H_\mathrm{dl}H_{x}}{2{H_k}^2}\cos(2\omega t).
\end{aligned}
\end{equation}

Analogously one can derive the Fourier expansion if the magnetic field is applied along the $y-$direction. In this case, the general form can then be written as

\begin{equation}
\label{eq:B8}
\begin{aligned}
V_{xy}(t) = &\mp\dfrac{I_0\Delta R_{AHE}H_\mathrm{fl}H_{y}}{2{H_k}^2}\\
&\mp \dfrac{I_0\Delta R_{AHE} \left(3H_\mathrm{fl}^2-8H_k^2+4H_{y}^2\right)}{8H_k^2}\sin(\omega t)\\
&\pm \dfrac{I_0\Delta R_{AHE}H_\mathrm{fl}H_{y}}{2{H_k}^2}\cos(2\omega t).
\end{aligned}
\end{equation}

Considering both initial magnetization states, as well as the different applied magnetic fields a general equation of $V_{xy}(t)$ can be given:

\begin{equation}
\label{eq:B9}
\begin{aligned}
V_{xy}(t) = &\dfrac{I_0\Delta R_{AHE}\left(H_\mathrm{dl}H_{x}\mp H_\mathrm{fl}H_{y} \right)}{2{H_k}^2}\\
&\mp \dfrac{I_0\Delta R_{AHE} \left(3(H_\mathrm{dl}^2+H_\mathrm{fl}^2)-16H_k^2+4(H_{x}^2+H_{y}^2)\right)}{8H_k^2}\sin(\omega t)\\
&\pm \dfrac{I_0\Delta R_{AHE}\left(H_\mathrm{dl}H_{x}\mp H_\mathrm{fl}H_{y} \right)}{2{H_k}^2}\cos(2\omega t).
\end{aligned}
\end{equation}

\section*{Acknowlegements}
The authors acknowledge funding from Senstronic Project. S.K.,S.Z.,F.B., C.A., J.S. and D.S. acknowledge funding from Österreichische Forschungsförderungsgesellschaft (FFG)  under the Project  Senstronic. C.A acknowledges funding from Austrian Science Fund (FWF) under Project No. P 34671. Computational results have been achieved, in part, by using the Vienna Scientific Cluster. C.S., R.G. G.J., M.K. acknowledge funding from Deutsche Forschung Gemeinschaft (CRC TRR 173 Spin+X, projects A01 and B02).
 
\section*{AUTHOR DECLARATIONS}
\subsection*{Conflict of interest}
The authors have no conflict of interest to disclose.

\subsection*{Intellectual property}
{A.S., and  D.S.} have a licensed Patent under the Paten number uS 17/220,129 \cite{suess_device_2021}.

\subsection*{Author Contributions}
S.K., F.B., C.A., D.S. wrote and improved micromagnetic code. S.K.,J.S. and D.S. derived analytical solutions, S.K. performed all computations and macrospin simulations. S.Z. assisted in all SOT related evaluations. C.S., R.G., G.J., M.K. performed the higher harmonics measurements, Mi.K., K.P. fabricated the samples. J.G., A.S., M.A., G.J., M.K., and D.S. supervised the project. S.K. prepared and wrote the initial manuscript. All co-authors have contributed to the final manuscript.
\subsection*{Data availability}
The data that support the findings of this study are available upon reasonable request from the corresponding author.

\end{document}